\documentclass[twocolumn,trackchanges]{aastex701}
\usepackage{amsmath}
\usepackage{graphicx}% Include figure files
\usepackage{dcolumn}% Align table columns on decimal point
\usepackage{bm}% bold math
\usepackage{xcolor}
\usepackage{soul}
\usepackage{comment}

\begin{document}

\title{Particle acceleration in Alfv\'enic turbulence with a strong guide field}

\correspondingauthor{Daniel Humphrey}
\email{dahumphrey@wisc.edu}

\author[orcid=0009-0002-2889-1493]{Daniel Humphrey}
\affiliation{Department of Physics, University of Wisconsin at Madison, Madison, Wisconsin 53706, USA}
\email{}

\author[orcid=0000-0001-6252-5169]{Stanislav Boldyrev}
\affiliation{Department of Physics, University of Wisconsin at Madison, Madison, Wisconsin 53706, USA}
\affiliation{Center for Space Plasma Physics, Space Science Institute, Boulder, Colorado 80301, USA}
\email{}

\author[orcid=0000-0003-1745-7587]{Vadim Roytershteyn}
\affiliation{Center for Space Plasma Physics, Space Science Institute, Boulder, Colorado 80301, USA}
\affiliation{Los Alamos National Laboratory, Los Alamos, New Mexico 87544, USA}
\email{}

\begin{abstract}
Magnetically dominated Alfv\'enic turbulence creates an effective environment for particle acceleration. However, when a strong mean field is present, traditional mechanisms like mirror and curvature acceleration become inefficient at explaining non-thermal particle energy distributions. Based on numerical and phenomenological study, we propose that in such turbulence, particles are accelerated in charge-starved current sheets, corresponding to current velocities approaching the speed of light. The distributions of the electric currents, plasma density, fluctuations of electric charge, as well as the energy distributions of accelerated particles, approximately follow log-normal statistics. Non-thermal particle distributions thus arise from particle acceleration in these charge-starved current sheets rather than from conventional Fermi-type particle acceleration by turbulent eddies.

\end{abstract}




\section{Introduction}

Extreme astrophysical environments such as {pulsar wind nebulae}, jets from active galactic nuclei, and black hole accretion disks are expected to host collisionless plasmas, whose rest-mass energy may be greatly exceeded by the energy density of the magnetic field. These highly relativistic, magnetized plasmas may experience large-scale instabilities and turbulence, dissipating energy into thermal heating and particle acceleration. Energetic particles produced by these systems provide observable synchrotron radiation \cite[e.g.,][]{atoyan1996,meyer2010,abdo2011a,abdo2011b}. 

Numerical simulations indicate that collisionless, magnetically dominated Alfv\'enic turbulence is a promising means of producing non-thermal particle energy distributions \cite[e.g.,][]{zhdankin2017a,comisso2019,comisso2022,zhdankin2020,nattila2020,nattila2022,vega2022a,vega2023,sebastian2025curvature,grosselj2026,meringolo2026,mbarek2026}. However, such dynamics were predominantly studied in the weak mean field regime, where magnetic field fluctuations were comparable to or significantly exceeded the uniform magnetic field. Much less is understood about particle acceleration and energy conversion for plasmas immersed in a strong mean field. This case is not only important for, e.g., large-scale evolution of stellar magnetospheres, but may also be relevant for understanding turbulence and dissipation at kinetic scales. Indeed, even in the absence of a uniform magnetic field, large scale astrophysical turbulence is expected to produce a hierarchy of eddies that each provide a mean field for the smaller scale turbulence ensconced within it, potentially leading to a relatively large local guide field near the kinetic inertial scales. 

Previous numerical studies of magnetically dominated Alfvenic turbulence in a pair plasma with a strong guide field \cite[e.g.,][]{chernoglazov2021,nattila2022,vega2022b,vega2024} have revealed several important differences compared to the weak-guide-field case. First, the energy spectrum of turbulent fluctuations at kinetic scales—scales smaller than the electron inertial scale—is significantly less steep when a guide field is present. This means that a larger amount of magnetic energy is concentrated at the gyro-scales of the particles. Second, the non-thermal acceleration of particles behaves differently in the presence of the guide field. Instead of following a power-law distribution, the probability density function of the accelerated particles resembles a lognormal form.

In this letter, we investigate turbulence and particle acceleration in highly magnetized, relativistic pair plasmas with both moderate and strong guide fields, utilizing particle-in-cell (PIC) simulations. We find that in both cases, fluctuations in electric charge, electric current, and plasma density are intermittent and exhibit a log-normal tail. However, the log-normal intermittency is more pronounced in the limit of a strong guide field, which may be associated with the smaller gyroradii of accelerated particles and sharper spatial contrasts in the intermittent fluctuations. {We argue that a significant role in non-thermal particle energization in this case is played by strong, charge-starved current sheets.} 

{Current sheets and magnetic reconnection events are known to form ubiquitously in magnetic turbulent flows driven at scales much larger than the plasma microscale (inertial and gyro scales). These phenomena affect Alfvénic turbulence by modifying the energy cascade at small scales and contributing to the dissipation of turbulent energy \cite[e.g.,][]{matthaeus_turbulent_1986,carbone1990, loureiro2017a,mallet2017a,mallet2017,loureiro2018,vech2018,walker2018,dong2018,loureiro2019,dong2022,vega2023b}. They also generally play a role in accelerating particles to relativistic energies, representing the initial stage of acceleration, known as the injection phase. After the injection phase, particles are accelerated to much higher ultrarelativistic energies through interactions with turbulent eddies \cite[e.g.,][]{zhdankin2017a,comisso2019,comisso2022,zhdankin2020,vega2022a,vega2023,vega2024,vega2025}. This acceleration arguably occurs due to magnetic curvature and mirror effects \cite[e.g.,][]{lemoine2024,vega2024b,sebastian2025curvature,das2025,boldyrev2026particleacc}, eventually reaching saturation when the gyroscale of the relativistic particles becomes comparable to the size of large-scale turbulent fluctuations.}

{We argue that this mechanism is significantly altered when a strong guide field is present. First, the efficiency of turbulent acceleration is suppressed because the curvature and mirror effects become much less effective in the presence of a strong guide field. Consequently, particle acceleration mainly occurs within current sheets. Second, perhaps counterintuitively, the acceleration within current sheets becomes more efficient when a strong guide field is present, compared to the scenario with a weak guide field.}

We connect the resulting particle energy distribution to magnetic-field and density intermittency via energy-balance arguments, similar to those recently discussed in \cite[][]{boldyrev2026particleacc}. We propose that non-thermal, log-normal distributions of energetic particles observed in magnetically dominated turbulence with a strong guide field arise from direct particle acceleration in charge-starved current sheets rather than from conventional Fermi-type particle acceleration by turbulent eddies.

\section{Numerical Setup}

We will analyze the results of two particle-in-cell simulations of decaying magnetically dominated turbulence in a pair plasma with the fully relativistic code VPIC~\cite[][]{bowers2008}. The simulations are performed in a ``2.5D" geometry, wherein the uniform guide magnetic field, $\bm{B}_0$, is applied in the $z$-direction, while the magnetic and electric fluctuations are varied only in the $x$-$y$ plane. 
This corresponds to a three-dimensional system with continuous translational symmetry along $z$. 

The code evolves all three vector components of the electromagnetic fields and particle momenta. Such a setup is expected to capture essential nonlinear dynamics of plasmas with moderate to strong guide fields, producing field energy spectra and particle energy distributions similar to those obtained in fully 3D simulations,  \cite[see, e.g.,][]{zhdankin2017a, zhdankin2018c, comisso2018, comisso2019,nattila2022,vega2023}. This has the advantage of allowing us to employ substantially greater numerical resolution.

We use a doubly-periodic $L\times L$ square domain. The turbulent fluctuations are initialized by imposing randomly phased large-scale perturbations of the magnetic field,  
\begin{align}
\label{deltaB0}
\delta{\bm B}(\mathbf{x})=\sum_{\mathbf{k}}\delta B_\mathbf{k}\hat{\xi}_\mathbf{k}\cos(\mathbf{k}\cdot\mathbf{x}+\phi_\mathbf{k}),
\end{align}
where the unit polarization vectors are chosen to be normal to the background magnetic field, $\hat{\xi}_\mathbf{k}=\mathbf{k}\times {\bm B}_0/|\mathbf{k}\times{\bm B}_0|$, in order to initiate shear-Alfv\'en type fluctuations. The two-dimensional wave vectors of the modes, $\mathbf{k}=\{2\pi n_x/L,{2}\pi n_y/L\}$, are chosen within the interval $n_x,n_y=1,...,8$. All the modes in Eq.~(\ref{deltaB0}) have the same amplitudes $\delta B_{\mathbf{k}}$, but random phases~$\phi_\mathbf{k}$. The initial root-mean-square value of the perturbations is given by $\delta B_0={\langle |\delta{\bm B}(\mathbf{x})|^2 \rangle^{1/2} }$, where the average is done over the simulation domain.  The nominal outer scale of turbulence can then be defined as $l=L/8$. The corresponding time scale, $l/c$, where $c$ is the speed of light, is used to normalize time in the numerical results. 

Domain sizes, timesteps, and particles per cell for our two runs can be found in Table \ref{table}. Our physical lengths are reported in units of $d_e=c/\omega_{pe}$, the nonrelativistic electron inertial scale, where $\omega_{pe}=\sqrt{4\pi n_0e^2/m_e}$ is the nonrelativistic electron plasma frequency and $n_0$ the mean density of each species.   The relative strength of the guide field is $B_0/\delta B_0=10$ for Run I and $B_0/\delta B_0 = 1$ for Run II.

We define two plasma magnetization parameters based on the strengths of the guide field and magnetic fluctuations: 
\begin{align}
\sigma_0=\frac{B_0^2}{4\pi n_0 w_0 m_e c^2},\  \
\tilde{\sigma}_0=\frac{(\delta B_0)^2}{4\pi n_0 w_0 m_e c^2}, 
\label{sigma_tilde}
\end{align}
where $w_0 m_e c^2$ is the initial enthalpy per particle. The initial distributions of both the electrons and the positrons are chosen to have the isotropic Maxwell-J\"uttner form with the mildly relativistic temperature $\Theta_0=k_BT_e/m_ec^2=0.1$.  For such a distribution, the specific enthalpy is given by $w_0=K_3(1/\Theta_0)/K_2(1/\Theta_0)\approx 1.27$, 
where $K_\nu $ is the modified Bessel function of the second kind. 

\begin{table}[t!]
\vskip5mm
%\hskip-2.0cm
\centering
\begin{tabular}{c c c c c c} 
\hline
{Run} & size $(d_e^2)$ & \# of cells & $\omega_{pe}{\delta} t$  & \# ppc & $B_0/\delta B_0$ \\
\hline
I & $1600^2$ & $23552^2$ & $6.0\times10^{-3}$ & 200 & 10 \\ 
II & $2000^2$ & $16640^2$ & $2.1\times10^{-2}$ & 100 & 1 \\ 
\hline
\end{tabular}
 \caption{Parameters of the PIC runs. {Here, $d_e=c/\omega_{pe}$ is the nonrelativistic electron inertial scale, $\delta t$ is the numerical time step{, and \# ppc is the particles per cell per species.}}}
\label{table}
\end{table}

We also define the relativistic inertial scale in the electron-positron plasma as $d_{rel}=\sqrt{w_0 c^2/(2\omega_{pe}^2)}$. The initial fluctuation magnetization in both of our simulations is $\tilde{\sigma}_0=40$, corresponding to guide field magnetizations $\sigma_0 = 4000$ in Run I and $\sigma_0 = 40$ in Run II. 

As the initial magnetic perturbations evolve, the system generates not only Alfv\'en modes but also a small fraction of ordinary modes. The magnetic fluctuations of these ordinary modes are polarized in the $x-y$ plane, while the electric fluctuations are primarily oriented in the $z$-direction \cite[e.g.,][]{vega2024}. To reduce the proportion of ordinary modes, we initialize a compensating current in the $z$-direction. This current adds a very weak tail to the initial particle distribution (up to $\gamma \approx 4$) but ultimately helps to lessen the impact of the ordinary modes on the turbulent dynamics. More details on this procedure may be found in (\cite{vega2024, vega2025, humphrey2026}). 

Finally, we analyze our simulations at 36 light-crossing times for Run I and 6 light-crossing times for Run II, which corresponds to several eddy turnover times in each run. At these respective snapshots, the initial perturbations have relaxed and turbulence is well-developed.

\section{Charge Starvation and Intermittency}

Turbulent eddies in a magnetically dominated Alfv\'enic turbulence, ${\tilde \sigma}\gg 1$, can produce magnetic shears corresponding to the regime of charge starvation, where the field-parallel electric current would formally approach the maximum possible value, $\left|J\right| = 2en_0c$ \cite[e.g.,][]{thompson1998,sobacchi2020,nattila2022,demidov2025,boldyrev2025}. 

When velocities associated with the current approach the speed of light, particles gain substantial relativistic inertia and require greater energy in order to produce even small changes in velocity. {This impediment of the electric current causes the presence of the field-parallel displacement electric field, which enhances energy transfer to the particles. The energization of particles in the field-parallel direction increases their effective inertial scales, resulting in more dispersive Alfvénic fluctuations \cite[e.g.,][]{vega2024}. The strengthened parallel electric field and increased dispersion tend to relax the frozen-in condition and locally reduce the magnetic gradient that causes charge starvation.}

{This magnetic self-regulation ensures that at scales below the charge starvation barrier, the magnetic gradients do not increase, and the spectrum of magnetic fluctuations remains steeper than~$k^{-3}$ \cite[e.g.,][]{vega2024b,boldyrev2025}. Additionally, this self-regulation process leads to a local increase in particle density (e.g., \cite{nattila2022}), which enhances the electric current in that area and helps mitigate the limitation caused by charge starvation.}

{An interesting question arises regarding whether a spontaneously generated charge-starvation layer can produce a field-parallel displacement current that exceeds the relativistic limit of $2en_0c$. We believe this question is related to the potential for generating electromagnetic waves in the nonlinear Alfv\'enic dynamics \cite[e.g.,][]{cairns2018,dokgo2019,yao2022}. Currently, there is a lack of analytical support for this possibility in magnetically dominated Alfv\'enic turbulence, and our numerical simulations have not demonstrated such behavior either. Consequently, we define the charge-starvation behavior in Alfv\'enic turbulence as a condition where the current necessary to support the magnetic gradient becomes relativistic, which imposes a nonlinear relativistic constraint on the field-parallel particle drifts.}

Both of our conducted runs exhibit the same fluctuation magnetization ${\tilde \sigma}$. Therefore, the likelihood of encountering charge-starving magnetic fluctuations is expected to be similar in both cases. However, self-regulation processes appear to be significantly influenced by the strength of the guide field. Below, we demonstrate that, somewhat counterintuitively, the effects of charge starvation described earlier are more pronounced in the presence of a strong guide field.

In the case of a weaker mean field, the gyroradii of the particles are larger. Related drifts are more effective at transporting the plasma across magnetic field lines, and finite Larmor radius effects smooth out gradients more efficiently. Therefore, we can expect less contrasting density and current variations in charge-starved current sheets. Charge-starvation and associated density and current fragmentation then have a less significant impact on the dynamics and statistics of the small-scale turbulence. The distribution of accelerated particles in this regime is predominantly shaped by interactions with turbulence fluctuations rather than charge-starved current sheets. 

For strong guide fields, however, gyroradii are much smaller than the electron inertial scale~$d_e$. As particles are energized by the electric field associated with current sheets, their momenta along the magnetic field lines increase while their field-perpendicular gyroradii remain unchanged due to magnetic moment conservation. As a result, their gyroradius-dependent drifts and the curvature drift are strongly suppressed, and the plasma has fewer avenues by which to regulate charge starvation. Alignment of particles with the magnetic field makes it difficult for gradients to relax.  Strong mean field thus preserves the effects of charge starvation at kinetic scales in the ensuing turbulent dynamics. 

\begin{figure}[h!]
\centering
\includegraphics[width=1.\columnwidth]{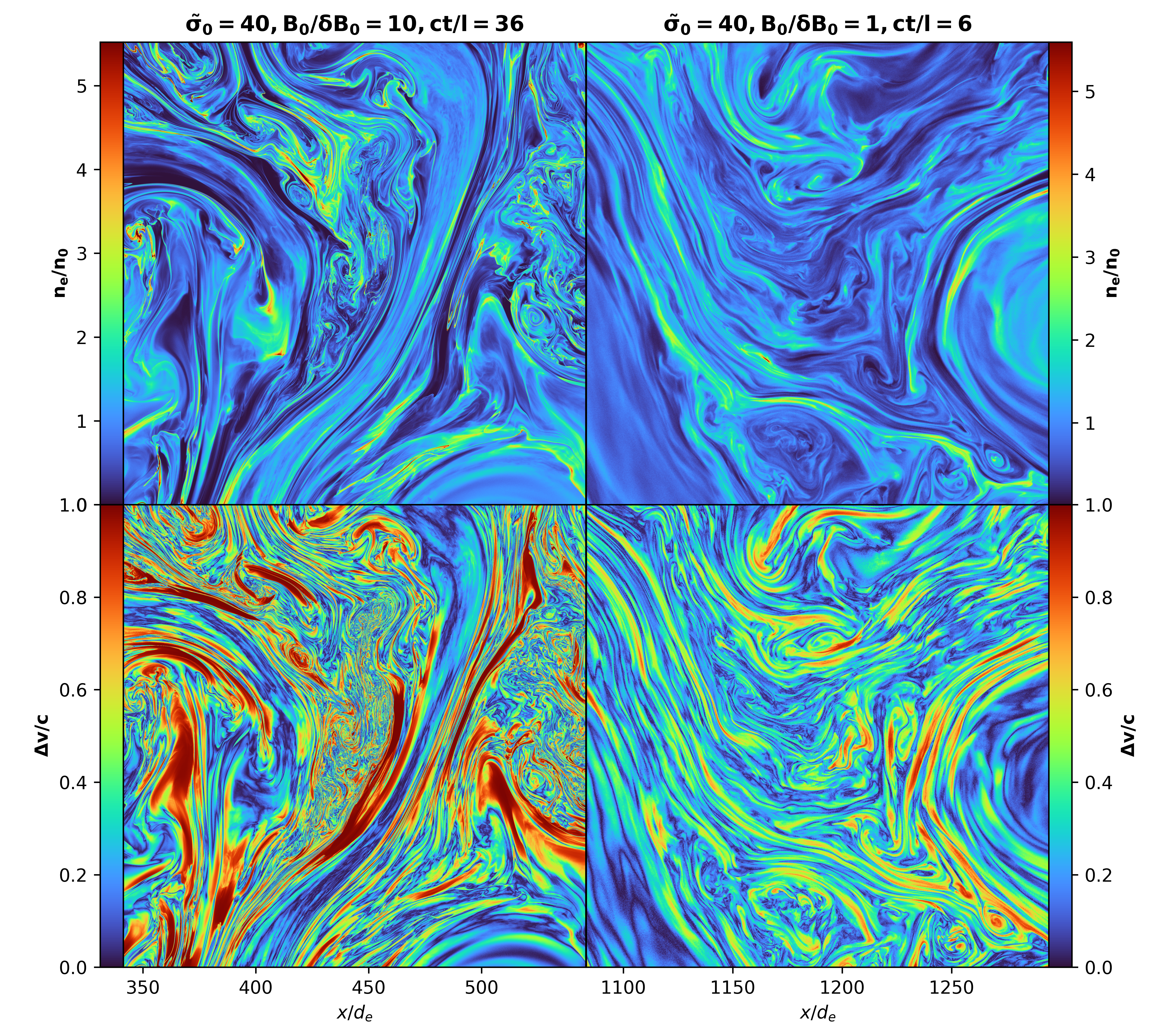}
\caption{On the left, the electron density $n_e$ and the velocity associated with the current $\Delta v = J/en$ are given in a portion of the simulation domain for the case of $B_0 /\delta B_0 = 10$. On the right, the same quantities are given for the case of $B_0 /\delta B_0 = 1$. Both regions are about $200 d_e$ in length, and were chosen to illustrate the multi-scale nature of turbulence. Though the electron density in the strong mean field case reaches higher values, we limit both colorbars to the same range to aid visual comparison. Note the sharper density gradients and richer small-scale structure of the turbulence in the case of a strong guide field, as well as the relative intensity and frequency of charge-starved regions (where $\Delta v/c\approx 1$).
\label{quad_ne_v}}
\end{figure}

\begin{figure}[h!]
\centering
\includegraphics[width=1.1\columnwidth]{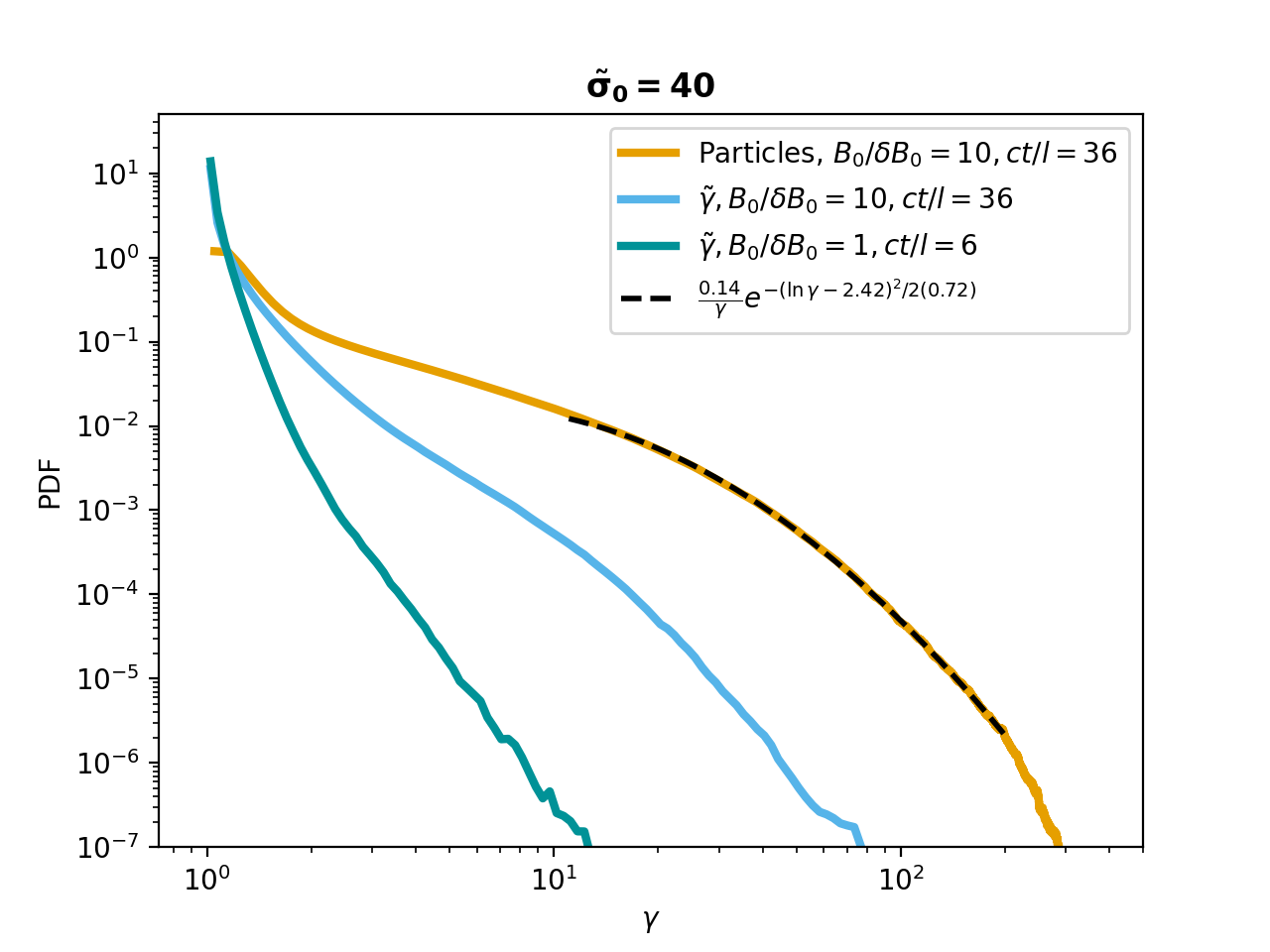}
\caption{$\gamma$ distribution for PIC particles in the case of $B_0 /\delta B_0 = 10$, as well as the bulk Lorentz factor associated with the current velocity $\tilde{\gamma} = 1/\sqrt{1 - \Delta v^2/c^2}$, with $\Delta v = J/en$, for both $B_0/\delta B_0 = 10$ and $B_0 /\delta B_0 = 1$. The tail of the particle distribution is fit by a log-normal. Severely charge-starved current sheets with $\tilde{\gamma} \gg 1$ are much more probable in the strong guide field case. 
\label{gamma_pdf}}
\end{figure}

Direct observation of the turbulence in both moderate and strong guide field regimes qualitatively supports this picture. Fig. \ref{quad_ne_v} shows two representative subdomains of a turbulent region, both about $200 d_e \times 200 d_e$ in size, and illustrates the electron density and the velocity associated with the current ($\Delta v = \left|J\right|/en$, where $n$ denotes total density) in each.\footnote{This expression corresponds to the {weighted average} of the current carrier velocities, defined on the interval $\Delta v/c \in [0, 1).$ This definition naturally excludes inertial motion, giving only the average bulk velocities contributing to the total current in the laboratory frame. If $\Delta v \approx c$, acceleration of particles cannot increase the current any further.} The simulation with a strong guide field, $B_0/\delta B_0 = 10$, contains more regions where the velocity associated with the current reaches close to the speed of light than the simulation with a moderate guide field, $B_0/\delta B_0 = 1$. In the case of a strong guide field, the density exhibits more filamented fine-scale structure, with regions of high density elongated into thin striations and accompanied by voids of low density. In the case of a moderate guide field, the overall morphology of the density is similar, but without the same small-scale behavior and extremes of compression and rarefaction. Rather, fluctuations at scales comparable to $d_e$ are smoothed over by the large gyroradii of particles.

We suggest that since turbulence with a strong guide field is less easily able to regulate charge-starvation, particles will experience the effects of the parallel electric field over longer periods of time. This will cause the Lorentz factor associated with the velocity of the current (that is, ${\tilde \gamma} = 1/\sqrt{1 - \Delta v^2/c^2}$) to grow larger.  This effect is seen in Fig. \ref{gamma_pdf}. In the case of a strong mean field, the bulk Lorentz factors associated with the current significantly exceed those found in the moderate mean field regime, demonstrating the presence of persistent charge-starvation. 

\begin{figure}[h!]
\centering
\includegraphics[width=1.1\columnwidth]{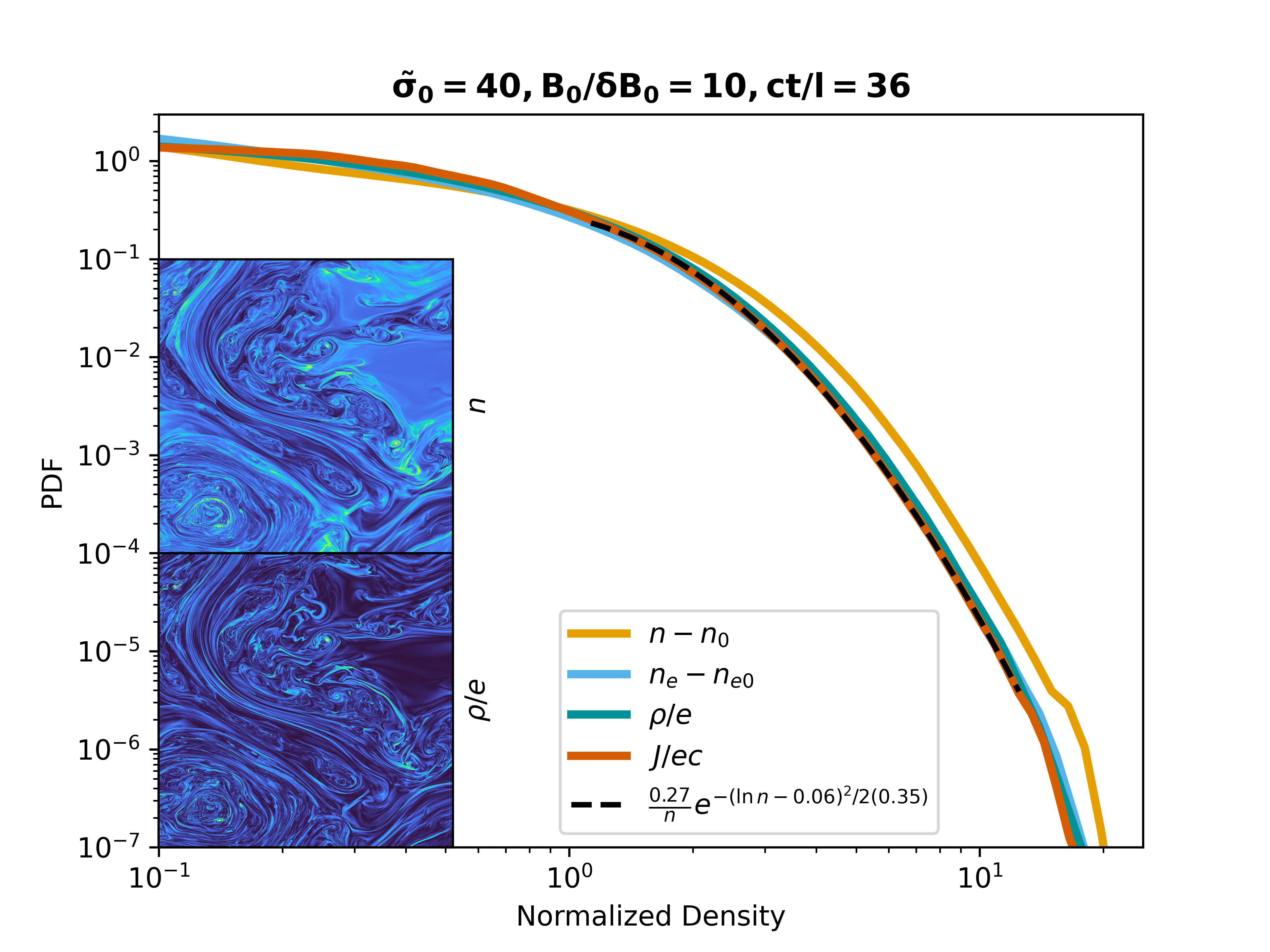}
\caption{ Distribution of total density $n = n_i + n_e$, $J/ec$, charge, and electron density $n_e$ for $B_0/\delta B_0 = 10$. The current is fit by a log-normal {(shown over the fitting interval)}. The total density and electron density are subtracted by their mean value{, and the absolute value of $\rho/e$ and $J/ec$ is used}. Each quantity has units of density and is normalized by the mean electron density. The visual structure of the density and charge are illustrated in the inset over a small portion of the simulation region. The log-normal tails are pronounced, and current is strongly coupled to density and charge.
\label{density_b10_pdf}}
\end{figure}

In nonrelativistic Alfv\'enic turbulence, magnetic fluctuations are known to be intermittent, with small-scale magnetic field variations as well as the electric current approximately following the log-normal statistics \cite[e.g.,][]{zhdankin2016}. We find similar log-normal intermittency in the relativistic case as well. The parallel current normalized by $ec$, the electron/positron density, and the charge all exhibit log-normal distributions (Fig. \ref{density_b10_pdf}).  The total plasma density fluctuations are also log-normal, but with a slightly broader tail. 

\begin{figure}[h!]
\centering
\includegraphics[width=1.1\columnwidth]{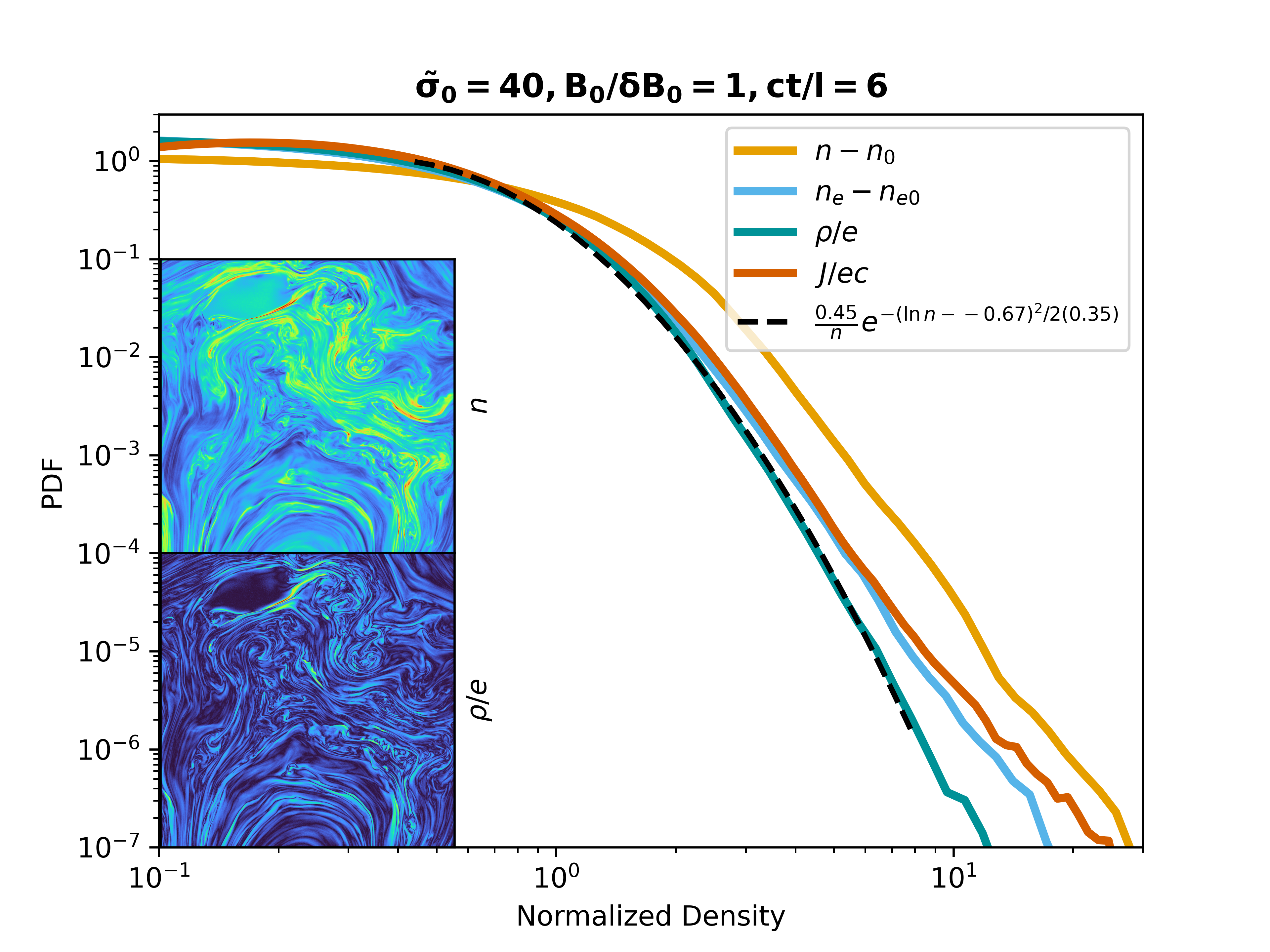}
\caption{ Distribution of total density $n = n_i + n_e$, $J/ec$, charge, and electron density $n_e$ for $B_0/\delta B_0 = 1$. The charge is fit by a log-normal {(shown over the fitting interval)}. The total density and electron density are subtracted by their mean value{, and the absolute value of $\rho/e$ and $J/ec$ is used}. Each quantity has units of density and is normalized by the mean electron density. The visual structure of the density and charge are illustrated in the inset over a small portion of the simulation region. The current is less strongly coupled to the density than in the case of $B_0/
\delta B_0 = 10$, and large fluctuations deviate somewhat from a log-normal form.
\label{density_b1_pdf}}
\end{figure}

Comparing Figures \ref{density_b10_pdf} and \ref{density_b1_pdf}, we find that magnetized turbulence with a strong guide field possesses density roughly as intermittent as the more naturally compressible moderate guide field regime. A more substantial difference is found in the charge $\rho/e$, which is more intermittent in the case of a strong guide field, consistent with the discussion in (\cite{vega2022b}). Both moderate and strong guide fields allow for charge imbalance in the field-parallel direction, but in the former case the plasma can more easily equalize the charge via field-perpendicular transport. In the latter case, large charge fluctuations are maintained in the intense layers of current formed by charge-starvation. 

\section{Particle Acceleration}

Magnetic curvature and magnetic mirrors have been found to play a key role in accelerating particles in strong Alfv\'enic turbulence \cite[e.g.,][]{lemoine2023b, vega2024b, das2025, sebastian2025curvature, boldyrev2026particleacc}. {Previously, it had been argued that curvature acceleration may yield a log-normal particle distribution function in the strong guide field case \cite[]{vega2024b}. However, the presence of a strong mean field greatly suppresses the efficiency of curvature and mirror acceleration. In \cite[]{vega2024b}, it is shown that the energy gain of relativistic particles by both curvature and mirror acceleration is proportional to an expression with $(\delta B_0/B_0)^2$. For the case of $\delta B_0/B_0 = 1/10$, then, the energy gain from these mechanisms will be small. Understanding the observed log-normal energization may, therefore, require a complementary mechanism that does not sharply decrease in efficacy with increased guide field. } 

\begin{figure}
    \centering
    \includegraphics[width=1.1\columnwidth]{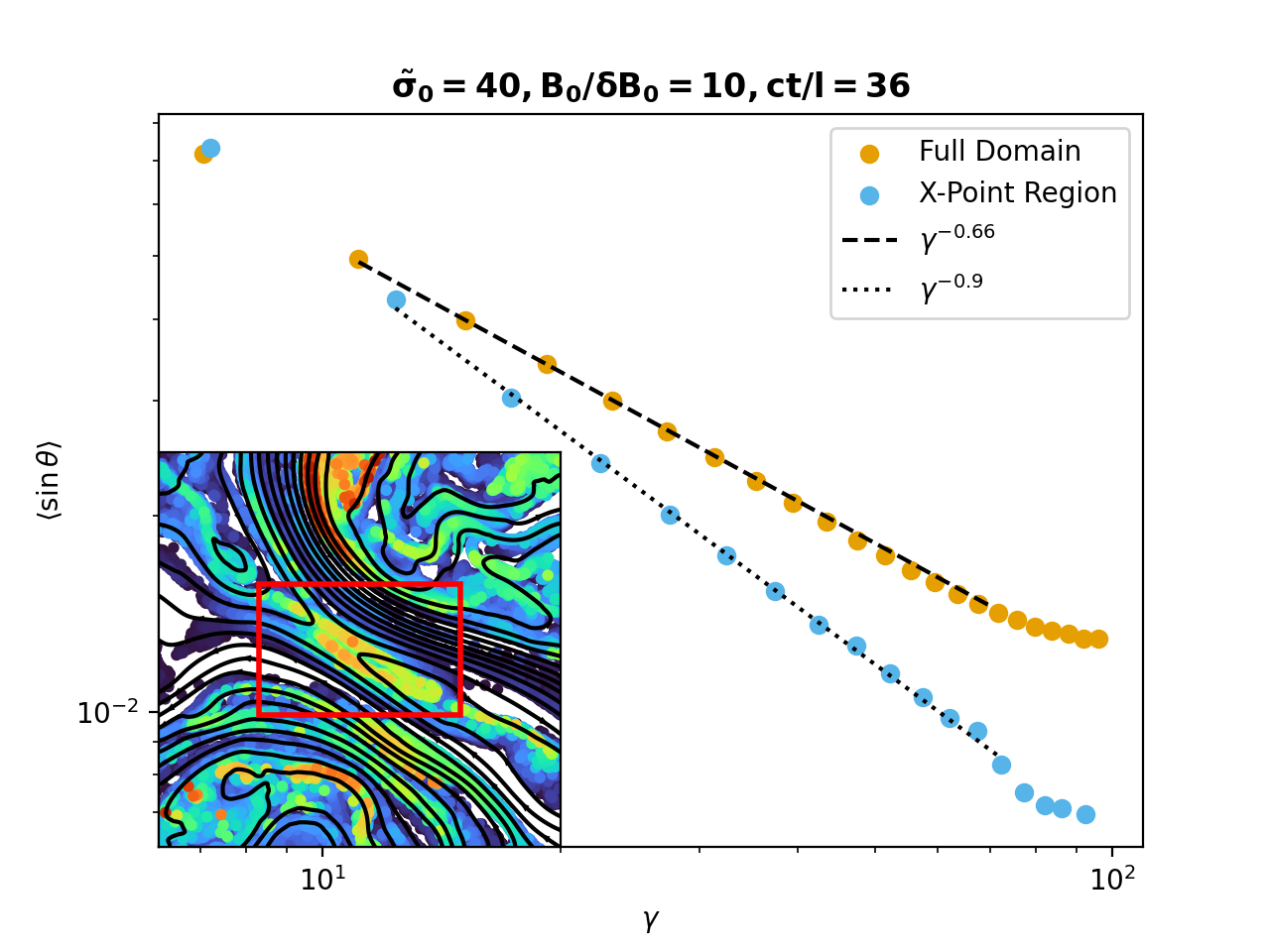}
    \caption{Average pitch angles as a function of energy shown for PIC particles over the whole simulation domain, as well as PIC particles taken \textit{only} from the region of active acceleration highlighted in the inset with a red box. The inset also shows magnetic field lines and particles colored by energy, ranging from blue ($\gamma \approx 5$) to red ($\gamma \approx 100$). $\gamma^{-0.66}$ and $\gamma^{-0.9}$ power laws are included for reference. Measurements are made via fully relativistic Lorentz transformations to the $E\times B$ frame of the plasma. Particles in the process of acceleration have pitch-angles declining very close to $\gamma^{-1}$.} 
    \label{pic_rcx_tracing_pitch}
\end{figure}

The presence of charge-starved current sheets in magnetically dominated turbulence may resolve this difficulty. When the speed of current-carriers approaches~$c$, the magnetohydrodynamic shear-Alfv\'enic cascade is unable to proceed, and the energy must be deposited into particles rather than smaller-scale Alfv\'enic fluctuations. Due to the conservation of magnetic moment below the inner scale of turbulence, $\sim d_e$, pitch-angles of accelerated particles are expected to decline as $\gamma^{-1}$, thus retaining the same perpendicular gyroradius (e.g., \cite[]{vega2024, vega2025, humphrey2026}), see also the discussions in~{\cite[][]{sobacchi2019,sobacchi2020,comisso2020,nattila2022,sobacchi2023}}.  Fig.~\ref{pic_rcx_tracing_pitch} shows average pitch angles as a function of $\gamma$ for both the entire domain and in a {particular current sheet X-point} where particles are being accelerated, showing good agreement with this prediction. High-energy particles therefore remain confined to the same spatial scale at which they are accelerated. 

\begin{figure}[h!]
\centering
\includegraphics[width=1.1\columnwidth]{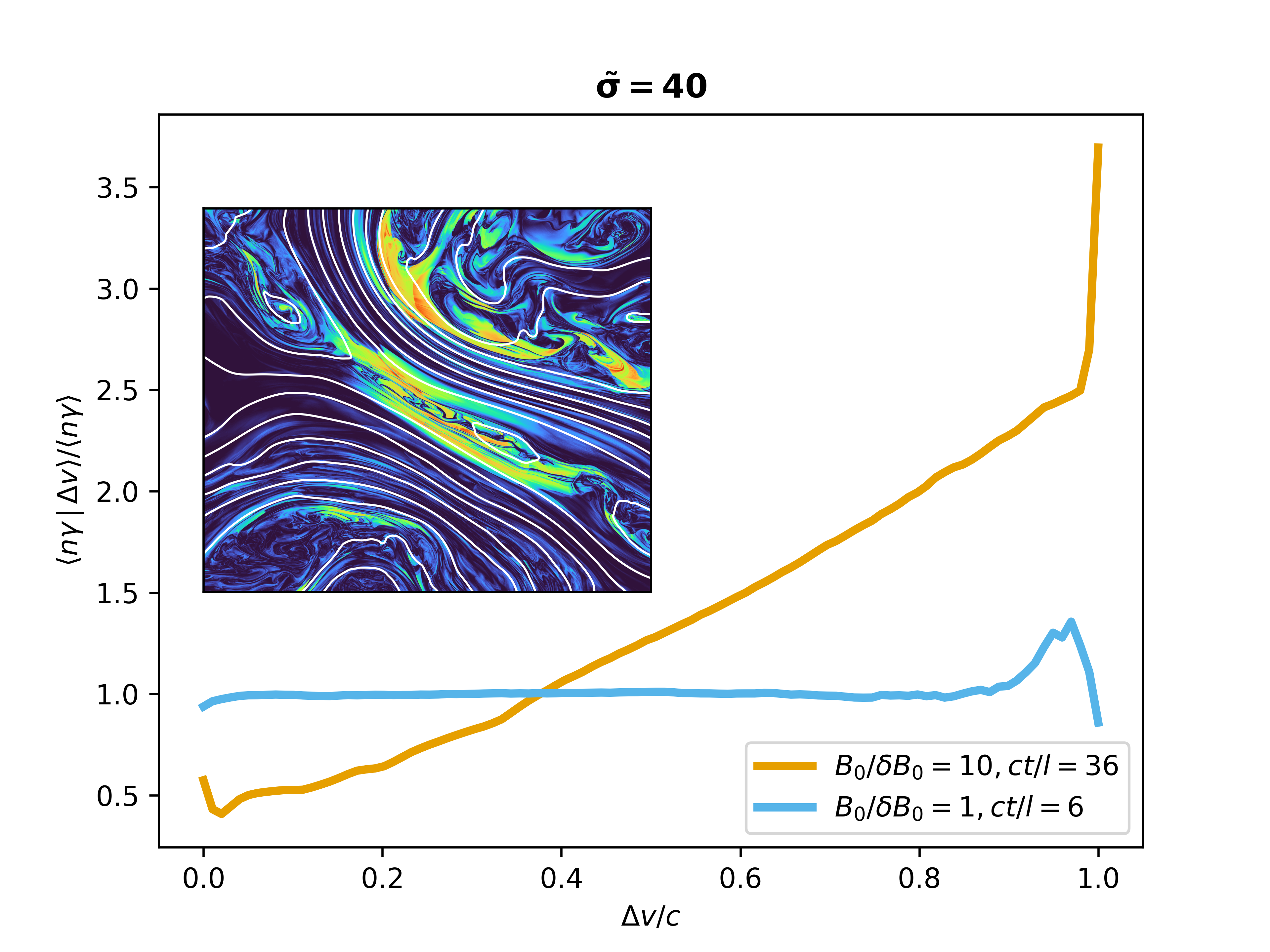}
\caption{The normalized, conditional expectation value of particle energy density as a function of bulk current velocity $\langle n\gamma | \Delta v\rangle /\langle n \gamma \rangle$ is shown for both $B_0/\delta B_0 = 10$ and $B_0/\delta B_0 =1$, where $\Delta v = J/en$. The energy density is calculated directly from PIC particles with $\gamma > 5$ taken at a particular timestep. The mean energy density is nearly uncorrelated with $\Delta v$ for the weak guide field case, but is strongly correlated for strong guide fields. For the case $B_0/\delta B_0 = 10$, $\tilde{\gamma} = 1/\sqrt{1 - \Delta v^2/c^2}$ is shown in the inset for the same region introduced in Fig. \ref{pic_rcx_tracing_pitch}, with a logarithmically scaled colormap from blue ($\tilde{\gamma} \approx 1$) to red ($\tilde{\gamma} \approx 85$).}
\label{e_dv}
\end{figure}

{We further examine the instantaneous distributions of PIC particles with $\gamma > 5$ in Fig \ref{e_dv}, calculating the conditional expected value of particle energy density $n\gamma$ as a function of the bulk current velocity, $\Delta v=J/en$, normalized by mean energy (that is, $\langle n\gamma | \Delta v\rangle /\langle n \gamma\rangle$). The energy density $n\gamma$ is therefore measured directly from the tail of the PIC particle energy distribution function, whereas $\Delta v$ characterizes the fluid-scale behavior of the current. In the case of $B_0 /\delta B_0 = 1$, the conditional expected value is approximately unity, indicating that $\langle n\gamma | \Delta v\rangle \approx \langle n\gamma \rangle$. This is consistent with acceleration occurring via alternate mechanisms such as magnetic curvature, leading to little correlation with $\Delta v$. However, the expected value of the energetic particle energy density increases with $\Delta v$ for $B_0 /\delta B_0 = 10$, suggesting a strong correlation between particle acceleration and charge-starved regions. } 

To determine the energy distribution of accelerated particles, we use an energy balance argument~\cite[see also][]{boldyrev2026particleacc}. For that, we notice that, in a qualitative difference with the case of non-relativistic turbulence, the electron density in a charge-starved current sheet of scale $\lambda$ adjusts to the gradient of the magnetic field according to the Ampere law:
\begin{eqnarray}
\label{n_B}
\frac{\delta B_\lambda}{\lambda} \sim \frac{4\pi}{c}enc.   
\end{eqnarray}
As discussed above, the turbulent energy supplied to an eddy of field-perpendicular scale $\lambda$ is then transferred to particles propagating along magnetic field lines inside this structure, mediated by the parallel electric field from the displacement current.  Denoting the field-parallel scale of the structure by~$l_\|$ and the Alfvén speed by~$v_A$, the lifetime of the structure is $l_\|/v_A\sim l_\|/c$, which is comparable to the particle crossing time of the eddy.  

Particles accelerated within this eddy may then gain energy $\gamma nm_e c^2$ up to the limit given by the magnetic energy supplied by the turbulent cascade on the same time, $\delta B_\lambda^2/8\pi$. We then obtain the dynamical energy balance condition:
\begin{eqnarray}
\label{gamma_B}
 \frac{\delta B_\lambda^2}{8\pi}\sim   \gamma nm_e c^2. 
\end{eqnarray}
Combining Eqs.~(\ref{n_B}) and~(\ref{gamma_B}), we obtain:
\begin{eqnarray}
\label{gamma_eq}
    \gamma \sim \frac{1}{2}\frac{\lambda^2}{d_e^2}\frac{n}{n_0}.
\end{eqnarray}

Given that particles are accelerated in charge-starving structures occurring at scales larger than $d_e$, the log-normal distribution of the density leads to an even broader log-normal distribution in $\gamma$, consistent with Fig.~\ref{gamma_pdf}. The charge-starvation scale $\lambda$ itself may have a certain distribution, which would further broaden the $\gamma$ distribution of accelerated particles. Large, intermittent current sheets (such as the one observed in Fig. \ref{pic_rcx_tracing_pitch}) can be expected to disproportionately contribute to the far tail of the particle distribution. In general, turbulent regimes that reach charge-starvation well before $d_e$ may have greatly enhanced particle acceleration.

The energy balance argument admits an important physical interpretation if we observe that the dynamic balance condition~(\ref{gamma_eq}) implies that the local relativistic inertial scale, $ d_{rel} \sim \sqrt{\gamma \left({n_0}/{n}\right)}d_{e} $, becomes comparable to the scale of the structure, $ \lambda $. Due to strong dispersion, the magnetic energy spectrum becomes steeper than~$ k^{-3} $ at scales smaller than~$ d_{rel}$~\cite[e.g.,][]{zhdankin2017a,comisso2018, vega2022b, vega2024}. This demonstrates how the system adapts to the charge starvation barrier: it enables the magnetic energy to propagate to smaller scales, but with a spectrum corresponding to subcritical current fluctuations.

{Finally, particle energization by charge-starvation may shed some light on the origin of unexpectedly strong intermittency of density fluctuations in the strong guide field regime (Fig. \ref{density_b10_pdf}). The energy balance argument suggests that sparse regions of highly relativistic plasma energy are created by intermittent magnetic fluctuations that approach the charge-starvation limit. The resultant strong pressure gradients may then fragment the current, density, and charge in the field-parallel direction. This effect is complementary to the inherent log-normal statistics of current in Alfv\'enic turbulence \cite[e.g.,][]{zhdankin2016}, which in the non-relativistic case manifests in fluctuations of the bulk velocity of current carriers. In the magnetically dominated case, similar statistics may require fluctuations of density due to the charge-starvation constraint. In either case, charge-starvation shows itself to be essential to the fully-developed dynamics and statistics of the magnetically dominated, strong guide field regime of turbulence.}

\section{Discussion and Conclusion}

Magnetized Alfv\'enic turbulence is an effective means of generating power-law energy spectra of particles, especially in the weak guide field setting where curvature and mirror acceleration can be very efficient. The presence of a strong guide field suppresses these important mechanisms, but the resulting turbulence still exhibits non-thermal particle acceleration. Instead of a power law, the statistics of accelerated particles are approximately described by log-normal distributions.

In this letter, we found through numerical and phenomenological studies that particle acceleration in this regime results from electric fields associated with charge-starved current sheets, rather than from Fermi-type interactions with turbulent eddies. Our findings indicate that the effects of charge starvation are more efficient in turbulence with a strong guide field compared to a weak guide field. This occurs because a strong mean field restricts the plasma's ability to self-regulate charge starvation, which significantly affects the statistics and dynamics of the turbulence. We found that the intermittency of charge-starved current sheets gives rise to log-normal statistics in the fluctuations of the magnetic field, plasma density, electric charge, and electric current.

We propose that the distribution of accelerated particles can be modeled using energy balance arguments. In the presence of a strong guide field, particles are accelerated along the magnetic field lines, while their gyroradii in the perpendicular direction remain unchanged. 
When the magnetic energy cascade is impeded by charge starvation within the turbulent structure, the energy from magnetic fluctuations is converted into the energy of the accelerated particles. As a result, the particle energy density reaches a balance with the magnetic energy density within that same structure. This relationship links the statistics of intermittency with particle acceleration. This dynamic energy balance provides a phenomenological explanation for the log-normal energy distributions of accelerated particles observed in numerical simulations.

{Future work could involve verifying this mechanism in full 3D runs. However, the presence of a strong guide field suppresses parallel magnetic fluctuations via the ``critical balance" condition $\delta B_0 / B_0 \sim k_\parallel/k_\perp$ \cite{goldreich_toward_1995}. Our acceleration mechanism depends precisely on these strong perpendicular gradients that are well-resolved in 2.5D, making it likely that our results will extend naturally to 3D. Encouragingly, \cite{nattila2022} found charge-starvation effects to be significant in their full 3D runs, even at $B_0 /\delta B_0 = 4$.}

We conclude that the mechanism of nonthermal particle acceleration in the presence of a strong guiding field is qualitatively different from that in a weak guiding field. Our findings indicate that the dynamics at small scales in magnetically dominated turbulence are log-normally intermittent and highly effective at transferring energy to particles in regions where charge-starved magnetic fluctuations are generated. {This mechanism remains effective at very strong guide fields, even as curvature and mirror acceleration become less efficient.} These results could enhance our understanding of particle acceleration, turbulence, and energy dissipation in high-energy astrophysical systems with strong guide fields.

\begin{acknowledgments}
 We thank the anonymous reviewer for their helpful comments. D.H. acknowledges helpful conversations with Fan Guo and Xiaocan Li. This work was supported by the U.S. Department of Energy, Office of Science, Office of Fusion Energy Sciences under award number DE-SC0024362. The work of D.H. and S.B. was also supported by the University of Wisconsin-Madison, Office of the Vice Chancellor for Research, with funding from the Wisconsin Alumni Research Foundation.  V.R. was also partly supported by NASA grant 80NSSC21K1692. Computational resources were provided by the Texas Advanced Computing  Center (TACC) at the University of Texas at Austin and by the NASA High-End Computing (HEC) Program through the NASA Advanced Supercomputing (NAS) Division at Ames Research Center. This research also used resources of the National Energy Research Scientific Computing Center, a DOE Office of Science User Facility supported by the Office of Science of the U.S. Department of Energy under Contract No. DE-AC02-05CH11231 using NERSC awards FES-ERCAP0028833 and FES-ERCAP0033257. 
\end{acknowledgments}

\newpage

\bibliography{references}{}
\bibliographystyle{aasjournalv7}



\end{document}